# Quenching statistics in Si and Ge SPADs using particle Monte Carlo simulation

**Philippe Dollfus[1,*], Jérôme Saint-Martin[1,2], Rémi Helleboid[1], Thibauld Cazimajou[1,§], Alessandro Pilotto[1,4], Arnaud Bournel[1], Denis Rideau[2] and Marco Pala[1,4]**

[1] Université Paris-Saclay, CNRS, Centre for Nanoscience and Nanotechnology (C2N), UMR 9001, 91120 Palaiseau, France
[2] Université Paris-Saclay, ENS Paris-Saclay, CNRS, SATIE, 91190, Gif-sur-Yvette, France
[3] STMicroelectronics, 38920 Crolles, France
[4] DPIA, University of Udine, 33100 Udine, Italy
[§] now with Université Claude Bernard Lyon 1, INL, UMR5270, CNRS, INSA Lyon, Ecole Centrale de Lyon, CPE Lyon, 69622 Villeurbanne, France
[*] philippe.dollfus@cnrs.fr

## ABSTRACT

Si- and Ge-based single-photon-avalanche-diodes (SPAD) are investigated by using self-consistent 3D Monte Carlo simulation, in a mixed-mode approach including the presence of a passive quenching circuit. This approach of transport allows us to capture all stochastic features of carrier transport and SPAD operation, not only for the avalanche triggering but also for the quenching process. Beyond the comparison of Si and Ge devices, we show in particular the strong inverse correlation between avalanche and quenching probabilities when tuning the bias voltage, which highlights the importance to find a tradeoff between these two probabilities for the optimization of SPAD operation.

## Introduction

Due to their high sensitivity and picosecond temporal response, single-photon-avalanche-diode (SPAD) photo-detectors have become key elements for many applications based on Light Detection And Ranging (LIDAR) principle that may be found in common usage systems like advanced driver assistance in automobiles [1]. Among many other applications, like quantum cryptography [2] and fluorescence lifetime imaging in biology [3], they are expected to play a crucial role in next-generation mobility systems such as self-driving cars, robots and drones, or for 3D imaging applications in smartphones.

The market of these devices is dominated by Si-based SPADs that can be integrated at low cost with standard CMOS process to produce large and ultra-sensitive detector arrays. However, they operate at wavelengths below 1 µm and there is a need of using lower bandgap materials to extend the spectral range of SPADs to short-wave infrared region where the background of solar radiation is reduced [4], the atmospheric transmission is enhanced [5] and the eye-safety threshold strongly increases. Germanium is a good candidate since it is sensitive to wavelengths of up to 1.6 µm and can be integrated with Si CMOS technology. That is why we investigate and compare here Si and Ge SPADs in the same basic configuration. The study is focused on the electrical behavior of devices from the generation of electron-hole (e-h) pair induced by photon absorption to the quenching of the avalanche process, by means of numerical simulation. The mechanism of light absorption is not described here.

SPAD operation has been already investigated experimentally on pure Ge photodiodes [6], but most of the recent proposals were based on Ge/Si heterostructures [7,8,9]. In this case, the Ge layer is used as absorber, while Si is used as multiplication layer.

Here, we propose a numerical comparison of simple designs of SPAD based on either pure Ge or pure Si, i.e., where both photon absorption and carrier multiplication take place in the same material. We have particularly focused this work on the quenching stage of operation that is not systematically analyzed for the optimization of this type of photodetectors. Many numerical studies were indeed focused on the early avalanche and detection stages of operation, e.g. using either TCAD [10][11] or particle Monte Carlo simulation of the Boltzmann equation [12,13,14], which is obviously relevant but certainly not sufficient to fully assess the SPAD performance. In terms of computational investigation of device design, the full SPAD operation in the presence of quenching circuit has mainly been discussed so far on the basis of compact modeling [15,16,17,18,19]. Computationally efficient Verilog-A model and TCAD mixed-mode simulation can be used as well [20]. In terms of physics of transport, though it is computationally demanding, the most accurate method is certainly the Ensemble Monte Carlo simulation that allows describing most of stochastic effects related to particle transport and ionization. Recently, by using an MC code self-consistently coupled to a 3D Poisson solver and able to include the passive quenching circuit within a mixed-mode approach, we have extended the capacity of this method to the physical time description of all stages of operation, from initial e-h pair generation to avalanche triggering and quenching [21,22], which makes possible to capture all statistical aspects of device operation. In particular, the notion of quenching probability appeared to be as important as those of avalanche breakdown probability and detection efficiency. We also showed that, even without the presence of traps, multiple avalanche current pulses may randomly occur during the quenching process when some particles remaining in the high-field region during the quenching stage are subject to new ionization events, which may trigger a new avalanche. This is consistent with recent experimental measurements on Si SPAD as a function of temperature [23].

In the present work, we not only compare the behavior of Si and Ge SPADs, but we also discuss the relationship between avalanche probability and quenching probability by using the supply voltage as a parameter. The inverse correlation between these two parameters may influence the optimization strategy of the device design and operation.

The article is organized as follows. In the second Section, we briefly describe the simulated devices and the main features of the MC model used. The third Section is dedicated to the results and their discussion, including the avalanche time and probability, and a specific focus on the quenching statistics with influence of quenching circuit elements and bias voltage. Conclusions are drawn in the final Section.

**Model and simulated devices**

The simulated structure consists in a Si or Ge PN junction of length $L$ = 1300 nm and square cross section of width $W$ = 300 nm (see Figure 1). It is connected in series with a passive $RC$ quenching circuit. The doping profile extends quasi linearly from $N_A = 6 \times 10^{17}$ cm$^{-3}$ to $N_D = 6.35 \times 10^{17}$ cm$^{-3}$ over a transition region of length 900 nm. A supply voltage $V_{BIAS}$ is applied between the terminals of the circuit composed by the SPAD and a ($R_Q, C_Q$) quenching circuit in series.

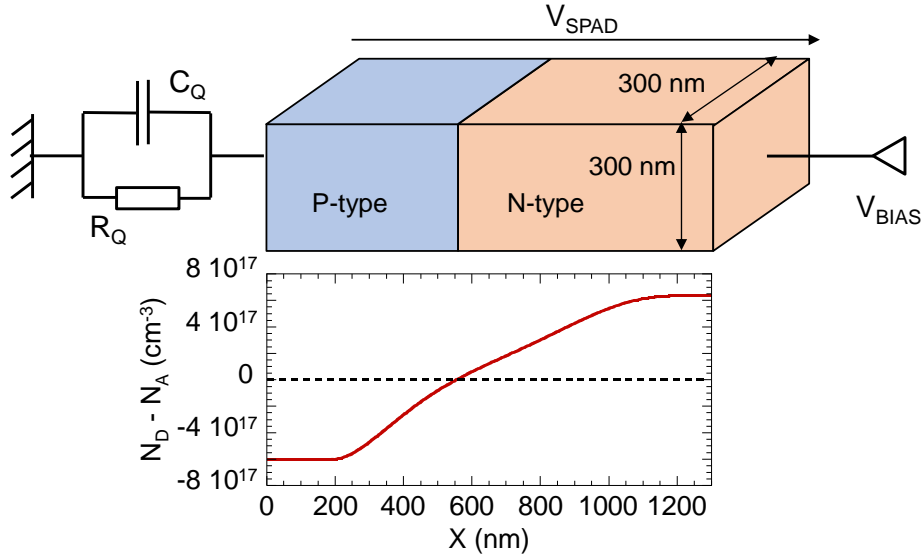

**Figure 1**. Doping profile and device cross section, including the external quenching circuit ($R_Q$, $C_Q$).

The 3D particle Monte Carlo code used is self-consistently coupled to the Poisson's equation [24]. In this model the analytic and non-parabolic conduction band of Ge is made of four ellipsoidal L valleys, i.e. eight half valleys in the first Brillouin zone, one spherical Γ valley and six ellipsoidal Δ valleys [25], while that of Si is made of six ellipsoidal Δ valleys and four isotropic L valleys [21]. For both Si and Ge the heavy and light hole bands are spherical and non-parabolic, with the room-temperature effective masses determined in [26]. The details on the parameters used for the calculation of electron-phonon scattering rates may be found in previous works for Ge [25] and Si [21]. To describe impact ionization processes, we implemented Keldysh like formulations of ionization rates [27] with parameters adjusted to fit the experimental field dependence of ionization coefficients for electrons and holes in Si [21] and Ge [22].

Compared to a real device, the cross section area of the simulated devices has been limited to a square cross section of 300 nm-width to make it compatible with self-consistent MC simulation in terms of computational time and resources but it does not change the main features of device operation, except of course the jitter that is not studied here. We just had to adjust the range of quench parameters $R_Q$ and $C_Q$ to higher and smaller values, respectively, than for a real device.

The connection between the SPAD and the passive ($R_Q, C_Q$) quenching circuit is made in a mixed-mode approach, as described in [21]. At each time step of Poisson's equation solution the SPAD voltage $V_{SPAD}$ is updated according to the current in the device (calculated using the Ramo-Shockley theorem [28]) and the quenching circuit elements. The initial single photon absorption is simulated by generating the resulting electron-hole (e-h) pair with energy and momentum randomly selected assuming thermal equilibrium distribution.

## Results and discussion

Prior to the analysis of SPAD operation, it is useful to extract some internal parameters as the capacitance and the resistance, which can be done from avalanche current- and charge-voltage characteristics of the device without quenching circuit. All simulations were made at room temperature. We have obtained $C_{SPAD}$ = 30.9 aF (20.8 aF) and $R_{SPAD}$ = 23.5 kΩ (29.8 kΩ) for Ge (Si) device [21,22]. From the avalanche current-voltage characteristics, we have also deduced the avalanche breakdown voltage $V_{BD}$ = 8.6 V and $V_{BD}$ = 15.3 V for Ge and Si devices, respectively. This difference, due to the smaller bandgap in Ge (0.76 eV) than in Si (1.12 eV), allows us to use a smaller bias voltage $V_{BIAS}$ in Ge devices, i.e. typically in the range 9.5-12 V instead of 17-20 V for the Si counterparts. This should lead to lower energy consumption in Ge SPADs than in Si SPADs, in spite of the risk of higher dark count rate that may require to operate at relatively low temperature [9]. In what follows, unless otherwise stated (in particular in the last sub-section), the results were obtained for $V_{BIAS}$ = 11 V and 18 V for Ge and Si SPADs, respectively.

**Avalanche Time and Probability**

The stochastic nature of the avalanche breakdown in SPAD, the very first stage of device operation, is well known [11]. It leads to a large spreading of the avalanche time $T_A$ and to the concept of avalanche probability $P_A$. It is illustrated here in Fig. 2, where we plot the cumulated number of impact ionization (II) events as a function of time for 20 random cases of photon absorption at the same position (center of the depleted region) but with random energy-wave vector coordinates of the initial electron-hole (e-h) pair. The time required to trigger the avalanche varies on a range of more than 15 ps, which may be obviously much larger if we consider the possibility to absorb the photon at any position in the device. We see in this figure (dashed lines) three cases where after a few II events the avalanche is not triggered. We can even observe cases without any ionization (not shown).

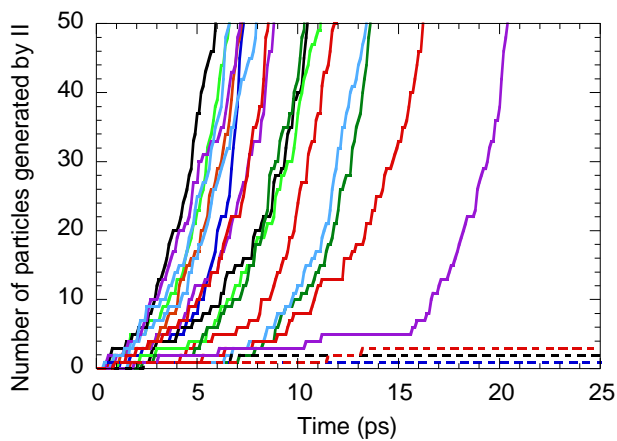

**Figure 2**. Twenty examples of avalanche triggering in Ge SPAD through time dependence of the cumulated number of particles generated by impact ionization after creation of an e-h pair at time $t$ = 0 by photon absorption in the center of the depleted region ($V_{BIAS}$ = 11 V). Each line corresponds to a random set of energy-wave vector coordinates of the initial e-h pair. The dashed lines represent cases where a few impact ionization events do not lead to any avalanche triggering.

It should be mentioned that these results for the early stage of the avalanche process have been obtained by including the self-consistence with Poisson's equation in the simulation, though it was not strictly necessary. The electrostatics does not change significantly during this short period of time and similar results can be obtained under frozen field, even with a simplified transport model [29]. In terms of avalanche time, the histogram reported in Ref. 22 has the form of a peak centered on 10.5 ps with a full-width at half maximum (FWHM) of about 6 ps, which is not incompatible with the experimental results obtained for an Si device of much larger size uniformly illuminated, with a 7.8 ps FWHM centered on 50 ps [30].

The resulting avalanche probability and average avalanche time are plotted in Fig. 3 as a function of $V_{BIAS} - V_{BD}$ for both devices. The avalanche time is defined here as the time needed after photon absorption for the avalanche current to reach the value $I_A$ = 5 µA, which will be discussed later with the help of Fig. 4. It is not surprising to see here that when increasing $V_{BIAS}$ the probability $P_A$ increases while the average avalanche time $T_A$ reduces, as a consequence of higher electric field in the depleted region of the device.

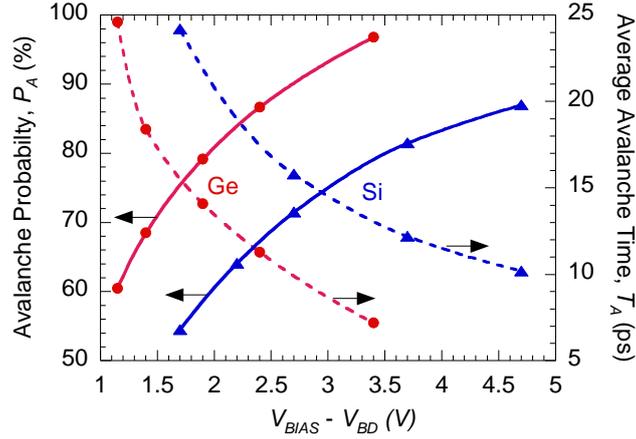

**Figure 3**. Avalanche probability (solid lines) and average avalanche time (dashed lines) as a function of excess bias voltage in Ge (red circles) and Si (blue triangles) SPADs.

**Quenching Statistics - Influence of Quenching Circuit Parameters**

It has been shown that the probabilistic character of SPAD operation is not limited to the avalanche triggering but also appears in the quenching process [21]. To illustrate this important feature, we plot in Fig. 4 for the Si device three typical cases of time response of the intrinsic SPAD voltage (Fig. 4(a)) and the current through the SPAD (Fig. 4(b)) after generation of an electron/hole (e/h) pair at time $t = 0$ in the center of the device. The three simulations correspond to three different random selections of the energy and wave vector coordinates of the initial e/h pair. The quenching circuit elements are here $R_Q$ = 600 kΩ and $C_Q$ = 0.1 fF with $V_{BIAS}$ = 18 V. It should be noted that for more realistic (higher) active area, the values of $R_Q$ and $C_Q$ should be approximately reduced and increased, respectively, in proportion to the area increase.

Starting from $V_{SPAD} = V_{BIAS}$, we observe here either a single avalanche followed by the quenching with return to initial state (red), a triple-avalanche with quenching (green), or no quenching at all (blue) with a final $V_{SPAD}$ close to $V_{BD}$. In all cases the voltage first falls abruptly below the breakdown voltage (more than 1 V below), and remains lower than $V_{BD}$ for a while, which tends to quench the avalanche process. It confirms what has been previously observed using an electric model including a stochastic description of the avalanche multiplication [31]. The return to equilibrium is reached within a time that obviously depends on the parameters $R_Q$ and $C_Q$ that control the time of discharge of the quenching capacitance $C_Q$, which may influence the probability of successful quench, as will be discussed later. If some carriers are still remaining in the depleted region of the device during the quenching process, i.e. when the voltage is re-increasing towards $V_{BIAS}$, they can trigger a secondary or even more avalanches, independently of any trapping effect (not included here), before finally quenching (green curve, triple avalanche) or may eventually prevent the quenching (blue curve). In the latter case, a kind of quasi-equilibrium state is reached with a voltage oscillating around the breakdown voltage $V_{BD}$. In this state it is not impossible that the SPAD finally quenches after a long time (some nanoseconds or more), but it is unlikely.

In terms of current, the main avalanche peak that charges the quench capacitance is similar for the three cases. After analyzing many configurations of these devices (depending on the material, bias voltage, and quenching capacitance), we decided that it was relevant to empirically fix the threshold of the avalanche current at $I_A$ = 5 µA. It means that the avalanche time $T_A$ is defined as the time needed for the avalanche current to reach this value $I_A$. In some cases, the main avalanche peak of current may be followed by secondary peaks (triple avalanche, green curve) before the current vanishes close to zero. In case of no quench (blue curve), we observe several peaks of current that oscillates around $I_A$ without vanishing.

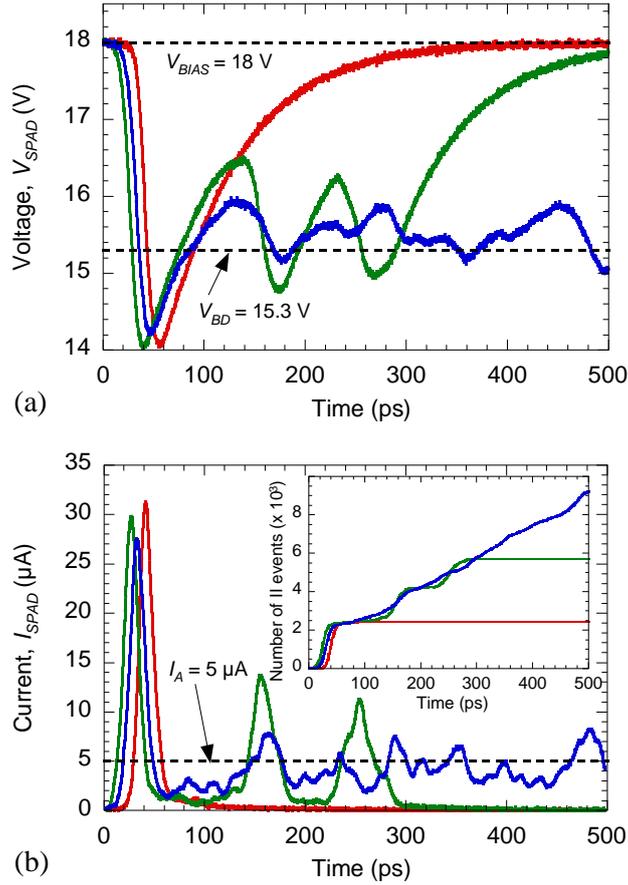

**Figure 4**. Time evolution of (a) voltage and (b) current in Si SPAD after generation at $t = 0$ of an initial e-h pair by photon absorption, with $R_Q = 600$ kΩ, $C_Q = 0.1$ fF, $V_{BIAS} = 18$ V. Each color corresponds to a random set of energy-wave vector coordinates of the initial e-h pair. The horizontal dashed lines represent the threshold avalanche current $I_A$, the bias voltage $V_{BIAS}$, and the breakdown voltage $V_{BD}$.

In the inset of Fig. 4(b) is plotted the cumulated number of ionization events experienced by electrons, the results being similar for holes with a lower level (not shown). Consistently with previous results, this number saturates in case of quenching (red and green lines) while it continuously increases in case of unsuccessful quench (blue).

This probabilistic character of SPAD operation deserves to be studied in more details as a function of the different parameters as the passive quenching circuit elements and the bias voltage. In what follows, we analyze the influence of the quenching resistance $R_Q$ and capacitance $C_Q$ on the quenching probability. Given the range of $R_Q$ and $C_Q$ values considered here, we have usually simulated a time window of 750 ps and a quench has been considered as successful if a time of 200 ps had elapsed since the last ionization event.

In Fig. 5(a), we plot the quenching probability $P_Q$ as a function of the resistance $R_Q$ for two values of $C_Q$. It appears that beyond a threshold resistance value that depends on $C_Q$, the probability $P_Q$ increases rapidly from 0 to 100%, both for Ge and Si SPADs. This behavior can be easily understood as a consequence of the increase of the time of discharge $\tau = R_Q C_Q$ of the capacitance that controls the time of quenching. It increases the chance to fully evacuate all excess carriers generated by the avalanche, and thus to make the triggering of new impact ionizations by remaining particles in the active region more difficult. In summary, increasing $R_Q$ at given $C_Q$ and $V_{BIAS}$, i.e. at a given amount of excess charges and given avalanche probability, just tends to enhance the quenching probability. This simple effect of $R_Q$ on the quenching has been for instance reported in [32] from electrical modelling.

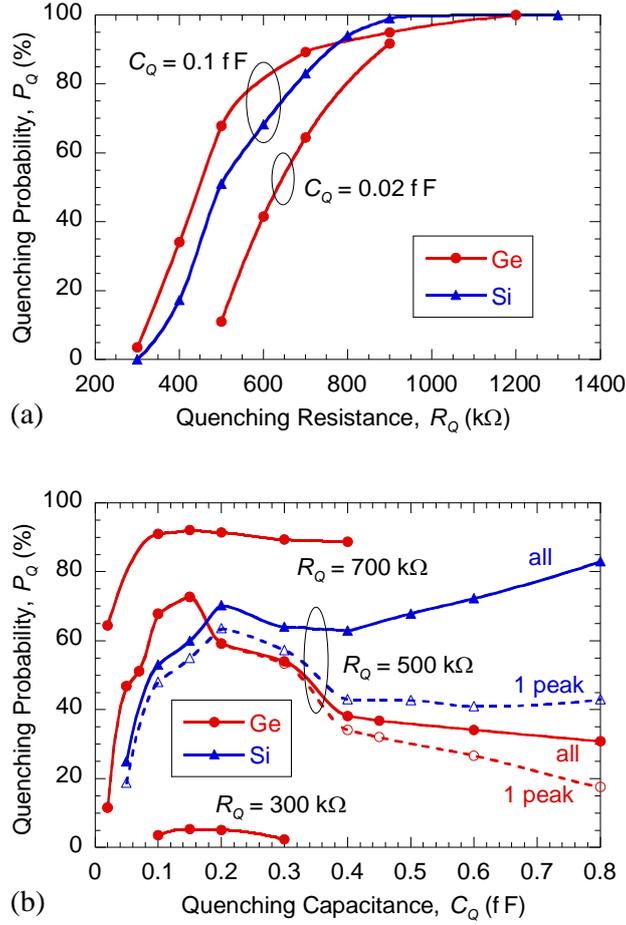

**Figure 5**. Quenching probability as a function of (a) the resistances $R_Q$ and (b) the capacitance $C_Q$, for both Ge SPAD (red lines, $V_{BIAS}$ = 11 V). Si SPAD (blue lines, $V_{BIAS}$ = 18 V). Solid lines: results considering all cases of successful quenching; dashed lines: results considering only cases of quenching after a single avalanche peak current.

The quenching probability as a function of the capacitance $C_Q$ is plotted in Fig. 5(b). We have distinguished here two different quenching probabilities: in solid lines, the probability $P_Q$ obtained by considering all cases of successful quenching, including after multiple avalanches, and, in dashed lines, the probability $P_Q^{1P}$ of quenching without error, i.e., obtained by considering only the cases of quenching after a single peak of current.

Let's consider first the probability $P_Q$ (solid lines). It appears that when varying $C_Q$ the competition between two antagonistic phenomena makes the effect on $P_Q$ less trivial than when varying $R_Q$. Indeed, on the one hand, the increase of $C_Q$ tends to increase the time of discharge $\tau$, which gives better chance to fully evacuate all excess charges and thus is likely to enhance $P_Q$, but, on the other hand, it also increases the charge stored in the capacitance, roughly equal to $C_Q \Delta V_{SPAD}$, and the amount of excess charges generated in the SPAD by impact ionization, which makes the quenching more difficult and thus reduces $P_Q$.

As a result of this competition, we observe in Fig. 5(b) that at low (high) $R_Q$ the probability $P_Q$ is low (high) and thus weakly dependent on $C_Q$, and that for intermediate resistance $R_Q$ = 500 kΩ the evolution of $P_Q$ as a function of $C_Q$ is quite erratic and not fully similar for Si and Ge devices. At low capacitance i.e., for $C_Q$ < 0.4 fF, the amount of excess charges generated by the avalanche is limited and the main effect of the increase of $C_Q$ is to enhance the time constant $\tau = R_Q C_Q$ that controls the quenching time, which increases the quenching probability. The probability behaves similarly for both devices with a maximum of about 70% in the range 0.15-0.2 fF. At higher $C_Q$ the higher amount of excess charges is more difficult to fully evacuate during the quenching process and increasing this capacitance makes the quenching more difficult. The quenching probability $P_Q$ tends to reduce for Ge but still slightly increases for Si.

This effect for high $C_Q$ is revealed in Fig. 6 where we plot in log scale the number of excess electrons in the Ge device as a function of time for two values of $C_Q$, i.e. $C_Q$ = 0.2 fF and 0.6 fF. We observe the avalanche peak of charges followed by the exponential decrease of charge according to the time constant $\tau = R_Q C_Q$. The avalanche peak is higher when $C_Q$ is higher and during the discharge of $C_Q$, i.e. during the quenching process, for $C_Q$ = 0.6 fF the number of excess electrons remains high (a few hundreds) for a quite long time (a few hundreds of ps). In the case of high avalanche probability (86.7% in this case) it results in a probability to have a secondary avalanches higher than for $C_Q$ = 0.2 fF, and thus in a smaller quenching probability, as shown in Fig. 5b.

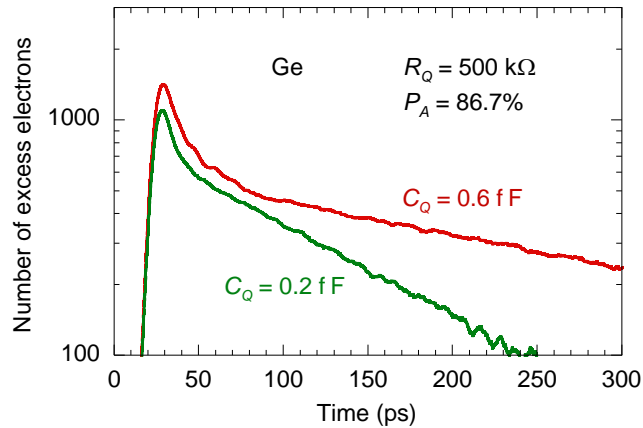

**Figure 6**. Number of excess electrons in the Ge device for two values of $C_Q$. Other parameters: $R_Q$ = 500 kΩ, $V_{BIAS}$ = 11 V, i.e. $P_A$ = 86.7%.

It is also interesting to see that the behavior of the probability of quenching without error $P_Q^{1P}$ (i.e., after one single avalanche, in dashed lines) is quite different. This probability is slightly higher for Si SPAD than for the Ge counterpart, but now the behavior is similar in both devices, close to that of $P_Q$ in Ge SPAD. A reasonable explanation may be found again in the difference of avalanche probability $P_A$ between both devices (Fig. 3). For the bias voltages considered here, $P_A$ is equal to 86.7% for Ge SPAD and only 71.5% for Si SPAD. Obviously, a higher avalanche probability makes the chance to trigger secondary avalanches higher and thus the quenching probability smaller. It is consistent with the lower value of $P_Q^{1P}$ for Ge than for Si. Additionally, a smaller $P_A$ increases the chance of successful quench after several secondary avalanches (secondary peaks of current), eventually after a quite long time of quenching, which explains the strong difference between $P_Q$ and $P_Q^{1P}$ in Si device at large $C_Q$, i.e., in the case of large number of excess carriers generated in the depleted region. A small value of avalanche probability is clearly an advantage to obtain a high quenching probability.

This effect is obviously strongly related to the bias voltage that controls the avalanche probability (see Fig. 3). For instance, increasing $V_{BIAS}$ for Si device increases the avalanche probability, which should tend to make the quenching behavior more similar to that of Ge device in Fig. 5b. Hence, it seems important to analyze the influence of the bias voltage on the overall performance of SPAD. This is the main object of the next sub-section.

**Quenching Statistics - Influence of Bias Voltage**

Actually, through the influence of bias voltage, we mainly intend to study here the influence of the avalanche probability $P_A$ for a given quenching circuit. We thus prefer to plot in Fig. 7 the quenching probability $P_Q$ as a function of $P_A$ instead of $V_{BIAS}$. The correspondence with $V_{BIAS}$ can be deduced from Fig. 3. For this figure, we have chosen an intermediate resistance $R_Q$ = 500 kΩ.

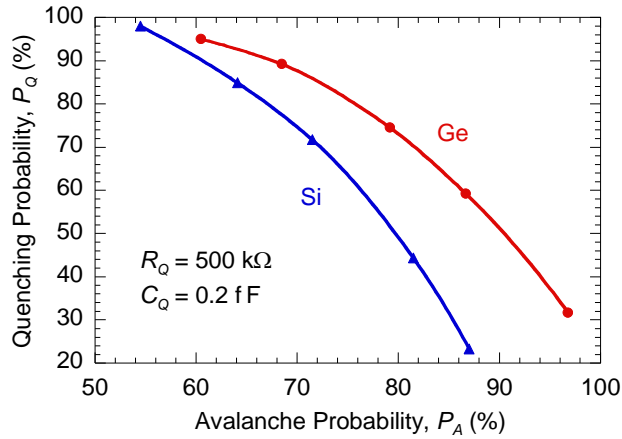

**Figure 7**. Quenching probability as a function of avalanche probability that is changed by tuning the bias voltage (see Fig. 2) in both devices for $R_Q$ = 500 kΩ and $C_Q$ = 0.2 fF.

The most remarkable result is that at given avalanche probability the quenching probability is higher in Ge SPAD than in Si device, which can be understood as follows. To make comparisons at a given probability, for instance at $P_A \approx$ 70 %, we have to consider $V_{BIAS}$ = 18 V for Si ($P_A$ = 71.5 %, see Fig. 3) and $V_{BIAS}$ = 10 V ($P_A$ = 68.5 %), which corresponds to an excess bias voltage $V_{BIAS} - V_{BD}$ of 2.7 V and 1.4 V for Si and Ge, respectively. At given $C_Q$, this smaller excess bias voltage in Ge leads to generate a smaller amount of excess charges, as shown in Fig. 8 where we plot the time evolution of the number of excess electrons in both devices. The smaller number of excess electrons (it is the same for holes, not shown) is thus easier to fully evacuate during the quenching process, which explains the higher quenching probability.

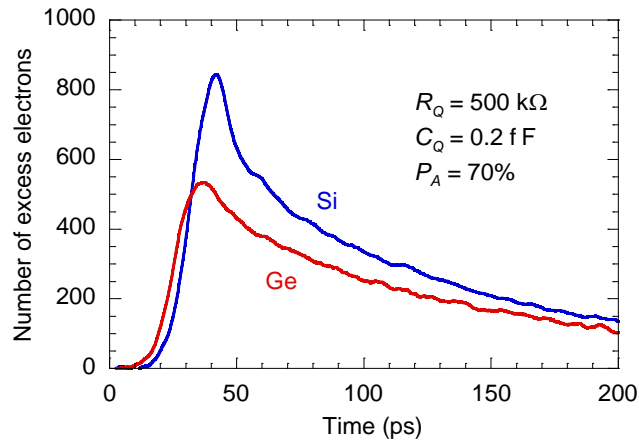

**Figure 8**. Typical time evolution of the number of excess electrons in both devices for an avalanche probability $P_A \approx$ 70% for given $R_Q$ and $C_Q$. $V_{BIAS} - V_{BD}$ = 2.7 V and 1.4 V for Si and Ge, respectively.

It is also very clear that a high avalanche probability is detrimental to the quenching probability $P_Q$ that can vary, in the case of Si, from 100% for $P_A$ = 54% ($V_{BIAS}$ = 17 V) to about 20% for $P_A$ = 87% ($V_{BIAS}$ = 20 V). We can thus anticipate that in view of a maximum detection efficiency, a trade-off has to be determined regarding the bias voltage, especially if it is not necessary to have a quenching probability strictly equal to 100%, i.e., if the circuit is designed with the possibility to regularly force the quenching of remaining unquenched SPADs.

It should be noted that by choosing a quenching resistor $R_Q$ of, for instance, 1 MΩ or more, this trade-off problem could be avoided since it would guarantee a high quenching probability. However, increasing the value of this parameter has a direct impact on the quenching time and thus on the minimum time between two possible detections (dead time) that is reflected in the so-called max count rate. If this metrics is critical for a given application, it may be relevant to try to reduce $R_Q$, which can make the trade-off problem quite crucial.

For such optimization we suggest to define a probability of "avalanche and subsequent quenching occurrence" $P_{AQ}$ that is just the probability that a single photon absorption triggers an avalanche multiplied by the probability to have a successful quench, that is $P_{AQ} = P_A \times P_Q$. If the avalanche is not triggered after photon absorption, the photon is not detected, and if the avalanche is not quenched, the device cannot detect more than one photon. This probability is thus relevant for practical device operation. However, it includes the possibility of secondary avalanches, i.e., secondary current peaks (sometimes called "secondary dark counts") that can be detected as a series of multiple photons though only a single photon was absorbed. Hence, it may be useful to also define a probability of avalanche and subsequent quenching occurrence without error as to consider only the quenching processes after a single avalanche peak of current.

In Fig. 9, we plot the two probabilities $P_{AQ}$ (solid lines) and $P_{AQ}^{1P}$ (dashed lines) as a function of $P_A$ for both devices. For $C_Q = 0.2$ fF (red lines for Ge, blue lines for Si), it is remarkable that the two probabilities have a maximum that is higher for Ge than for Si. At low $P_A$, the quenching probability is close to 100% so that reducing further the avalanche probability reduces the probabilities of avalanche and subsequent quenching. In contrast, at high $P_A$ a further increase of the avalanche probability is over-compensated by the decrease of the quenching probability, which reduces the probabilities $P_{AQ}$ and $P_{AQ}^{1P}$.

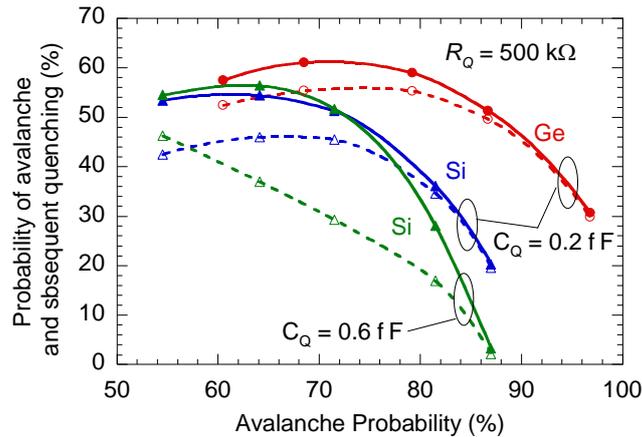

**Figure 9**. (Solid lines) Probability of avalanche and subsequent quenching occurrence $P_{AQ}$ and (dashed lines) $P_{AQ}^{1P}$ as a function of avalanche probability $P_A$ for both devices. $C_Q = 0.2$ fF and, only for Si device, $C_Q = 0.6$ fF. $R_Q = 500$ k$\Omega$.

At higher $C_Q = 0.6$ fF (green lines for Si), we observe a higher difference between $P_{AQ}$ (solid) and $P_{AQ}^{1P}$ (dashed), which is consistent with the difference between $P_Q$ and $P_Q^{1P}$ previously observed in Fig. 5(b). Additionally, these probabilities $P_{AQ}$ (solid) and $P_{AQ}^{1P}$ (dashed) fall more abruptly to zero when increasing the avalanche probability. Indeed, at high $C_Q$ the number of excess carriers generated in the depleted region by the avalanche process is high and increasing the bias voltage further increases this number of excess carriers, which makes more difficult to fully evacuate these carriers and thus to successfully quench the avalanche. It should be noted that for Si device at high avalanche probability of 87%, corresponding to $V_{BIAS} = 20$ V, the quenching probability decreases when $C_Q$ increases from 0.2 fF to 0.6 fF, which is fully consistent with what has been observed in Fig. 5b for the Ge device with a similar avalanche probability of 86.7%.

In terms of quenching and detection efficiency, it is thus useful to have a high resistance $R_Q$, since it may even guarantee to have a $P_Q$ of 100% but the capacitance $C_Q$ must be kept to reasonably low values to avoid quenching instabilities and errors in the photon detection, especially if the device technology suffers from possible variability issues.

## Conclusion

The avalanche probability is often considered as the most important parameter to optimize for a good detection efficiency. However, the present results highlight the fact that if the quenching is not artificially forced, the highest avalanche probability is not always the best option for SPAD optimization. Actually, according to the quenching circuit elements, when adjusting the supply voltage a trade-off may have to be found between avalanche and quenching probabilities to obtain the highest probability of "avalanche and subsequent quenching", which may depend on the targeted application.

It should be mentioned that this new metrics $P_{AQ}$ cannot replace the two probabilities $P_A$ and $P_Q$. It has however the advantage of emphasizing the antagonistic behavior of $P_A$ and $P_Q$ when tuning the bias voltage, which suggests that it may be useful to consider as an additional parameter when refining the optimization and/or the operation of a given SPAD design.

In terms of device performance, it appears that for a given avalanche probability, from the point of view of detection and quenching statistics, for this particular device structure the Ge SPAD provides a higher quenching probability than the Si counterpart, which also leads to a better probability of complete detection.

**Acknowledgements**

This work was supported in part by the French Industry Ministry ("Nano2022" and "MODSPADGE" projects under IPCEI program) and by the French ANR with the project GeSPAD (ANR-20-CE24-0004).

## Author Contributions

PD performed simulations of Ge SPAD, some simulations of Si SPAD and wrote the first draft of the manuscript. TC made some simulations of Si SPAD. PD, JSM, RH, TC, AP and MP developed the SPAD simulation methodology using MC method. PD, JSM and AB developed the core of the simulation code. All authors participated in the discussion and analysis of results and in the finalization of the manuscript. MP and DR supervised our overall SPAD simulation project.

## Competing interests

The authors declare no competing interests.

## Data availability

The datasets used and/or analyzed during the current study available from the corresponding author on reasonable request.